# Frequency Diverse Arrays (FDAs) vs. Phased Arrays: On the Application of FDAs for Secure Wireless Communications

M. Fartookzadeh

Department of Electrical and Computer Engineering, Malek Ashtar University, P. O. Box 1774-15875, Tehran, Iran.
Email: Mahdi.fartookzadeh@gmail.com
Submitted: 4/18/2020

**Abstract: Application of range-angle focusing frequency diverse array (FDA) for secure wireless communication has been proposed in recent years. However, it can be proven that the FDAs are not suitable for secure wireless communication and the same properties of the focusing FDA can be obtained by using traditional phased arrays, more effectively. Besides, in this paper, it is indicated that FDAs are not good candidates for constructing time-variant focusing beampatterns, while constructing time-invariant focusing beampatterns is impossible at all. Furthermore, some incorrect considerations about the FDAs are highlighted.**

**Keywords: frequency diverse array, focused beamforming, pulsed phased array.**

## I. INTRODUCTION

Range-angle focusing frequency diverse array for the purpose of secure wireless communication has been proposed in the recent years [1-3]. It is now well-known that it is not possible to construct an antenna array with a time-invariant range-focusing beampattern in the farfield [4, 5]. Also, it has been proven that the time variance of the beampattern produced by an arbitrary antenna array exactly follows its range variance, meaning that it is not possible to construct an array with time-invariant range-variant beampattern, in any way including the frequency diverse array (FDA) technique [6]. Nevertheless, the focusing point should be time-invariant for secure wireless communication to avoid transmitting from one point to other unwanted range-angle.

In this paper, the main purpose is to rectify the incorrect believe about the FDAs, which reflect that they can provide more secure communication link by using the focused beampatterns since the focusing point is time-variant. Furthermore, it is indicated that the time-variant focusing beampatterns can be produced by using pulsed phased arrays, more easily and efficiently.

The FDA is compared with the phased array in section II and it is indicated that the FDA is in fact nothing more than a phased array with time-variant phase differences. In section III, it is suggested that the FDA is not appropriate for secure wireless communication and the phased array can be more effective for this application.

## II. FDA VS. PHASED ARRAY

Linear FDAs and phased arrays are constructed from multiple numbers of antenna elements placed on a straight line. The difference is on the excitation signal which is with same frequency for the phased array and with tiny frequency shifts for the FDA. Schematics of an $N$-element linear FDA and phased array are depicted in Fig. 1 (a) and (b), respectively. The received signal, $S_r$, in the farfield region at the given range and time ($r = r_0$ and $t = t_0$) is obtained by using the superposition principle for both situations. Hence, $S_r$ of the antenna array with $N$ elements, each labeled by an integer number, $n$ ($n \in [0, N-1]$), is given by

$$S_r(t_0, r_0, \theta) = \sum_{n=0}^{N-1} \frac{k_n}{(r_n)^2} S_i^n \left(t_0 - \frac{r_n}{c}\right) = \sum_{n=0}^{N-1} \frac{k_n}{(r_0 - nd\sin\theta)^2} S_i^n \left(t_0 - \frac{r_0 - nd\sin\theta}{c}\right). \tag{1}$$

where $S_i^n$ is the excitation signal of the $n$-th element. $k_n$ can be divided into the transmitter part $k_t^n$, and the receiver part $k_r$, which is independent of $n$. Now by considering the farfield condition, $r_0 \gg (N-1)d$, and combining $k_t^n$ with $S_i^n$ as $S'^n_i$, the received signal will be

$$S_r(t_0, r_0, \theta) = \frac{k_r}{(r_0)^2} \sum_{n=0}^{N-1} S'^n_i \left(t_0 - \frac{r_0}{c} + \frac{nd\sin\theta}{c}\right). \tag{2}$$

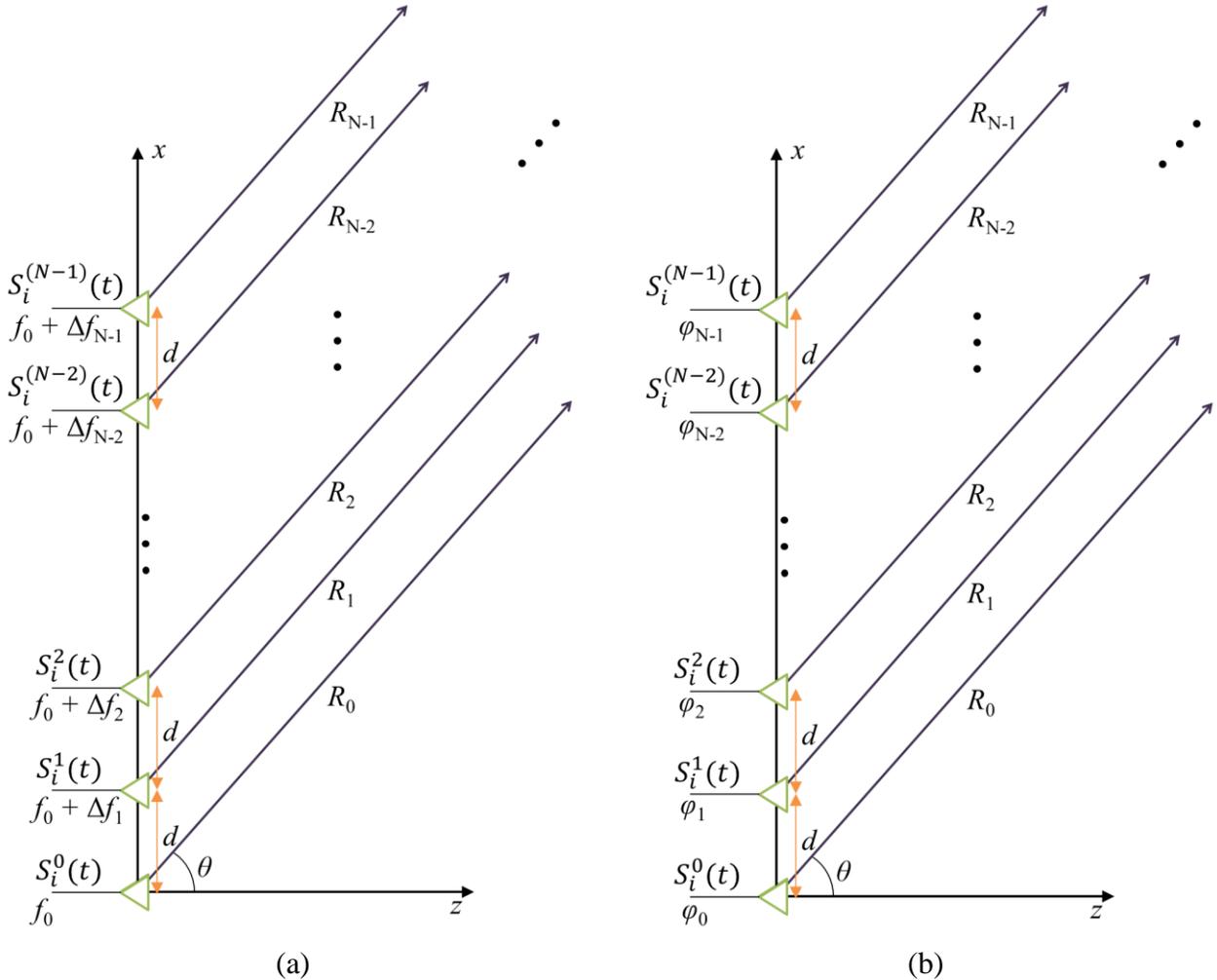

Fig. 1. Schematics of the N-element linear (a) FDA and (b) phased array.

The range dependency of the received signal is only the inverse-square $1/(r_0)^2$, and the $r_0/c$ delay from the initial signal at the array point [6].

In the analysis of antenna arrays the received signal is usually divided into two parts; element factor (EF) and array factor (AF), $S_r = AF.EF$. The AF represents the influence of array, without considering the effects of antenna elements. In addition, the common parts of the AF in the summation which do not have effects on the beampattern shape can be removed. Consequently, AF of the antenna array with N elements, is given by [4]

$$AF(t, r, \theta) = \sum_{n=1}^{N} w_n e^{j\left[2\pi f_0 \frac{nd\sin\theta}{c} + 2\pi\Delta f_n\left(t - \frac{r}{c} - \frac{nd\sin\theta}{c}\right) + \phi_n\right]}, \tag{3}$$

for both FDA and phased array situations. Assuming $Nd\Delta f_n \ll c$, the AF will be

$$AF(t, r, \theta) = \sum_{n=1}^{N} w_n e^{j\left[2\pi f_0 \frac{nd\sin\theta}{c} + 2\pi\Delta f_n\left(t - \frac{r}{c}\right) + \phi_n\right]}, \tag{4}$$

where $f_0$ is frequency of the first element and $(f_0 + \Delta f_n)$ is the frequency of $n$-th element ($\Delta f_n = 0$ for the phased array). $c$ is the speed of electromagnetic (EM) wave propagation and $d$ is the spacing between elements. $w_n$ and $\phi_n$ are the weight and phase of the $n$-th element, respectively.

It can be observed that there are three terms in the exponent of AF. The first is shared between the FDA and the phased array, the second is for the FDA and the third is for the phased array. The only difference is that the effect of a constant phase difference does not change with time, while the frequency difference is multiplied by $(t - r/c)$, leading to the time-range variance of the AF. Hence, there is nothing special for the FDA and the same AF can be obtained by using time-variant phased array (with time-variant phase differences between elements), leading to the time-range variance of the AF, similarly.

As a consequence, there is a general fact that FDA and phased array do not have principal distinctions. However, for some practical applications phased arrays might be better and for some applications FDAs might be more suitable, such as the proposed direction finding method by using FDAs in [7, 8]. Therefore, in the rest of this paper it is explained that FDAs are neither suitable for constructing focused beampatterns nor for secure wireless communication.

## III. Secure Wireless Communication by Using FDAs?

At the start of this section, it is noteworthy to observe the time dependency of the beampattern for a 15-element FDA in Fig. 2. The plot in Fig. 2 (a) is in fact repeated from Fig. 4 (b) of [1], to have a typical example without loss of generality. However, small differences can be observed between the sidelobes, which can be due to the small variations between the calculated Chebyshev coefficients. However, it does not change the method of time variation of the beampattern. Following the previously mentioned facts about constructing FDA beampatterns [4], the excitation should have been started earlier to make sure that it has enough time to reach the most faraway point in the plot. For example, it requires 0.5 ms to arrive at the point on 150 km distance. Hence, t = 0 is only a convention for the beampattern in Fig. 2 (a) and the excitation began earlier. The beampattern at t = 1 ms and t = 2 ms is also depicted in Fig. 2 (b) and (c), respectively. It can be observed that the beampattern as well as the focusing point is moving with the speed of light. Hence, it is not appropriate for secure communications. On the other hand, if we assume that this moving beampattern has an application, it will be indicated that the same beampattern can be constructed by using the phased array more easily and more efficiently.

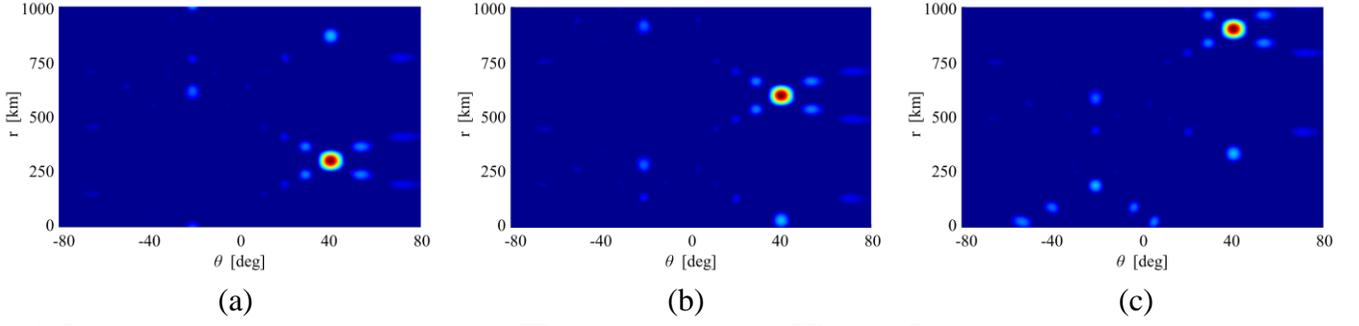

(a)           (b)           (c)

Fig. 2. Time-variance of the focusing point of the FDA; beampattern of the FDA with Chebyshev frequency offsets on the range-angle axes (a) at $t = 0$ [1], (b) at $t = 1$ ms and (c) $t = 2$ ms.

The key point is the relation between time and range dependency of the produced beampattern, $t - r/c$, which result in an equivalent beampattern on time and range dimensions, excepting the $c$ factor in the denominator. Therefore, the same beampatterns can be obtained by using a phased array excited by pulse functions as illustrated in Fig. 3. The first pulse in Fig. 3 (a) is a rectangular pulse with 0.27 ms pulse-width and the second pulse is a Gaussian pulse function with full width at half maximum (FWHM) of 0.35 ms or $\sigma = 0.15$ ms (FWHM $= 2.355\sigma$). These pulse functions are selected to obtain similar beampatterns as for the FDA in Fig. 2. However, the beampattern produced by the pulsed phased array is more controllable in the range dimension as will be explained, shortly.

Assuming a 15-element phased array with Chebyshev coefficients, the beampattern for a simple continues wave is as depicted in Fig. 4 (a) (same as the beampattern in Fig. 4 (a) of [1]). However, the beampattern will be as shown in Fig. 4 (b) at $t = 0$, if the array is excited with the rectangular pulse function in Fig. 3 (a). It can be observed that there is no sidelobe for this type of excitation and array. Especially, no signal is transmitted excepting at the 0.27 ms around $t = -1$ ms to be observed on the range dimension at $t = 0$ excepting on 81 km around the 300 km distance. Hence the phased array is indeed better for secure communications, since the undesired sidelobes are removed. The other advantage is the controllability of the focused point on the range dimension. For example, the width of the focusing point can be reduced to 27 km if the pulse-width is reduced to 0.9 ms and so on.

In addition, other pulse functions can be used instead of the rectangular pulse functions. For example, one can use the Gaussian pulse function as depicted in Fig. 3 (b) to produce a more similar beampattern with the FDA. Consequently, the beampattern will be as illustrated in Fig. 4 (c), which is very similar to Fig. 2 (a) without any sidelobes. The variation of the beampattern with time is similar to the FDA as indicated in Fig. 2.

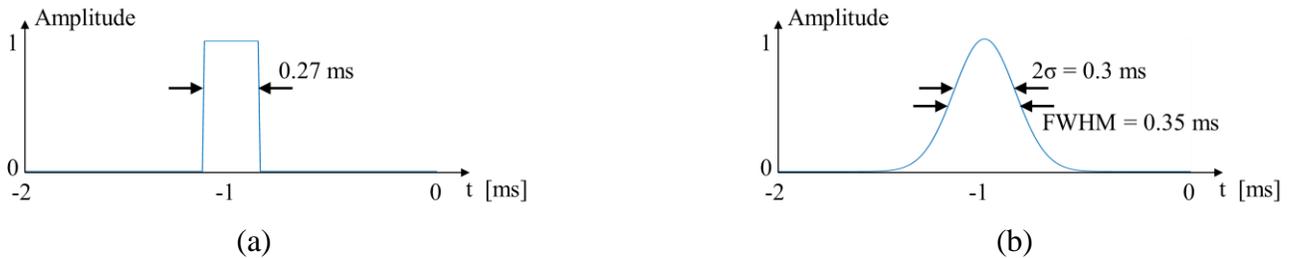

(a)           (b)

Fig. 3. Excitation pulse functions; (a) rectangular pulse and (b) Gaussian pulse.

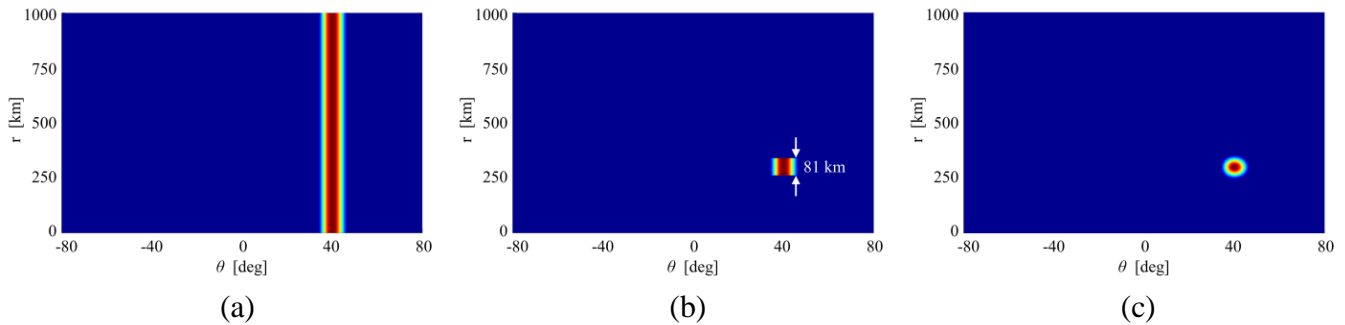

Fig. 4. Beampattern of the phased array on the range-angle axes; (a) continues wave [1], (b) rectangular pulse with 0.27 ms pulse-width at $t = 0$ and (c) Gaussian pulse with $\sigma = 0.15$ ms at $t = 0$.

Further than the explained disadvantages of the FDA compared with phased array, another important disadvantage might appear in realization of the FDAs. This practical drawback can occur in the input ports of antenna-elements of such focusing FDAs. The meaning of focusing beam pattern for FDAs is that the radiated power by the array elements are intensifying at an instance of time and diminishing in other times. In other words, the radiated EM waves by the antenna elements are constructive on a specified angle at a moment and destructive in other times. Hence, the radiated EM waves by the antenna elements might be available at other times on other undesired angles. Otherwise, where does the excited power go in the times which no EM wave is radiated? When the antennas are excited and no radiation occurs at any angle, it means that the power is returned to the inputs and thus the return loss is high. This is why the active reflection coefficient is crucial for evaluation of multi-port antennas as well as antenna arrays [9]. Consequently, it is indicated that an FDA is not a good candidate for constructing time-variant focusing beampattern and thus for secure communication, which is the main purpose of this paper.

## IV. Conclusions

In this paper, it is indicated that the recently introduced frequency diverse array (FDA) is indeed equivalent to a special type of the phased array with varying phases. In fact, the tiny frequency differences between the input signals of the array elements bring about continues variations of the phases of the signals, leading to beam direction changing in traditional FDAs. In addition, the so called focusing FDA is made of the EM waves which are constructive at a moment and destructive in other times. The resulting beampatterns are similar to the beampatterns of pulsed phased array, while in the phased array the focusing width on range axes can be adjusted, easily. As a consequence, it is more convenient to produce dot-shaped beampatterns by using pulsed phased arrays instead of FDAs.